\documentclass[12pt,preprint]{aastex}
\slugcomment{Accepted: \emph{Astronomical Journal}}

\newcommand{\be}{\begin{equation}}
\newcommand{\ee}{\end{equation}}

\begin{document}

\title{Further Constraints on the Presence of a Debris Disk in the Multiplanet System Gliese 876}
\author{P. D. Shankland\altaffilmark{1,2}, D. L. Blank\altaffilmark{1}, D. A. Boboltz\altaffilmark{2}, T. J. W. Lazio\altaffilmark{3}, \\
G. White\altaffilmark{1}}

\altaffiltext{1}{Centre for Astronomy, School of Mathematical \&
Physical Sciences, James Cook University, Townsville QLD 4811 AU
(paul.shankland@jcu.edu.au, david.blank@jcu.edu.au,
graeme.white@jcu.edu.au)}

\altaffiltext{2}{U. S. Naval Observatory, 3450 Massachusetts Ave NW,
Washington, D.C. USA 20392-5420 (paul.shankland@usno.navy.mil,
dboboltz@usno.navy.mil)}

\altaffiltext{3}{Naval Research Laboratory, 4555 Overlook Avenue,
SW, Washington, DC 20375-5351 (joseph.lazio@nrl.navy.mil)}

\begin{abstract}

Using both the Very Large Array (VLA) at 7mm wavelength, and the
Australia Telescope Compact Array (ATCA) at 3mm, we have searched
for microwave emission from from cool dust in the extrasolar
planetary system Gliese~876 (Gl~876). Having detected no emission
above our 3$\sigma$ detection threshold of 135 $\mu$~Jy, we rule out
any dust disk with either a mass greater than 0.0006 M$_{\oplus}$ or
less than $\sim{250}$~AU across. This result improves on previous
detection aperture thresholds an order of magnitude greater, and it
has some implications for the dynamical modeling of the system. It
also is consistent with the Greaves et al. hypothesis that relates
the presence of a debris disk to close-in planets. Due to the
dust-planetesimal relationship, our null result may also provide a
constraint on the population or composition of the dust and small
bodies around this nearby M dwarf.

\end{abstract}


\section{Introduction}

The M4 dwarf star Gl~876 harbors one of the nearest multiplanet
systems detected to date. At a Hipparcos-determined distance of 4.69
pc (Perryman et al. 1997), this star is orbited by three planets
(Delfosse et al. 1998; Marcy et al. 1998; Marcy et al. 2001).  The
outer two planets are gas-giants, while the innermost is likely to
be of terrestrial mass (Rivera et al. 2005). The semi-major axes of
these planets ranges from 0.02 to 0.2~AU.  We first targeted Gl~876
to detect any optical transits as described in Shankland et al.
(2006).

The search for and study of planetary systems and dust disks around
M dwarfs is a relatively new endeavor and the results to date have
not been entirely consistent.  Only a few debris disks are known
around M stars, and Gautier et al. (2007) found no new detections in
a \textit{Spitzer} search for dust disks around 123 late-type
dwarfs. However, the nearby M dwarf AU~Mic shows a well-resolved
debris disk, whose radius is between $50$ and $210$~AU (Kalas, Liu,
\& Matthews 2004). As well, Gl~842.2 (Lestrade et al. 2006) was
shown to have a $\sim{300}~AU$ disk, and Gl~182 (Liu et al. 2004)
was found to have a $\sim{120}$~AU one. These suggest that Gl~876
could reveal a disk if it had one.

A disk detection is important in that it would offer a better
understanding of the Gl~876 system, and it would begin to provide
clues about planet formation about M dwarfs. M~dwarfs comprise
three-fourths of the galactic population, making any detection
important to understanding plant formation for this most common
type. A detection would also help to characterize a particularly
diverse and nearby planetary system. Separate from the potential to
witness disk disturbances which might reveal planets, a disk
detection would further determine whether mature dust disks signal
the presence planets in a system, and help constrain whether debris
disks are the ubiquitous result of planetary formation. Conversely,
detecting \emph{no disk} would still be helpful as it would set the
system's upper mass limit.

Right now the inclination in Gl~876 is not well understood, and
assessing any such `tilt' would help indicate whether the
inclination approaches $i \approx 50^{\circ}$, or is more like $i
\approx 90^{\circ}$. The former is contended as a result of radial
velocity reductions, as discussed in Rivera et al. (2005) and
Shankland et al. (2006). The latter is derived using astrometric
data from Hubble Space Telescope's Fine Guidance Sensor (FGS) by
Benedict et al. (2002). Learning a system's inclination (presuming
the disk is coupled to the plane of orbits) is invaluable to a
complete understanding of systems and the true mass(es) of their
planets. Optical radial velocity measurements are limited to
providing a mass limit, M~($sin~i$). Any additional constraint on
$i$ helps constrain the M~($sin~i$) mass (and vice versa), which can
then lead to constraints on density, composition, and ultimately,
habitability.

Modern millimeter interferometers (e.g., VLA and ATCA) offer
arcsecond resolution which can give information beyond a simple
detection. Such capability could also shed qualitative light on a
system's dynamical inclination, as shown in Lestrade et al. (2006).
If the Gl~876 system were to contain a debris disk, the extent of
which exceeds just $\sim{5}$~AU (which is our resolving power at
4.69~pc), then imaging observations could not only resolve the disk
but also provide some geometric constraint on the inclination of the
system. Of course the disk would have to have some `optical
thickness' in order to observe any gradient consistent with an
inclination.

Lestrade et al. (2006) found a disk around GJ~842.2, and refer to
the one about Gl~182 found by Liu et al. (2004). Both disks are
large enough ($\sim$100~AU) that they could be cleanly resolved by
us. In their own quest for dust about our target Gl~876, Trilling,
Brown, \& Rivkin (2000) demonstrated their ability to discern
inclination from their dust disk observations of three \emph{other}
(G-type) stars using Cold Coronagraph with the Infrared Telescope
Facility (IRTF), an instrument with less resolving power than VLA
and ATCA. These precursor observations assured us that some general
inclination information could be extrapolated from the appearance of
any disk.

Other specific issues would be clarified with an improved
understanding of dust disks about M dwarfs. Since dust has a short
lifespan throughout a disk due to radiation pressure,
Poynting-Robertson drag and gravity, it has to be regenerated to
maintain the disk. This is done by the larger bodies which collide
and so shed dust. Th$\acute{e}$bault, Augeraeau, \& Beust (2003)
assert this for $\beta$ Pictoris, and such a disk about Gl~876 would
infer that a solar-like Kuiper-like belt might exist. While this
relationship is expected in many disks, the planetesimal-dust
relationship is not so clear for M dwarfs.

A lack of dust may also support the Greaves et al. (2004) hypothesis
-- that debris disks are uncommon around stars with close-in giants,
presumably due to the short lifetimes of the parents of debris disks
and a sweeping effect. For Greaves' hypotheses, Gl~876 is a good
test. A disk would also specifically imply that rocky or icy bodies
exist that are left over from the system's formation. A detection
would would also mean any planet-inferring asymmetries could be
studied, as done for the HR 4796 disk (Wyatt et al., 1999).
Certainly Gl~876's radial-velocity-derived 2:1 planetary resonance
between Jovian planets `b' and `c' would lead one to seek additional
correlations with any of dust disk, such as Quillen \& Thorndike
(2002) reported in $\epsilon$ Eridani's ring. A general census of M
star dust disks would generally improve undertanding of planet
formation in general, as described in Apai, Luhman \& Liu (2007).

Our search for thermal dust emission using the Australia Telescope
Compact Array at 3mm (94 GHz) and the Very Large Array at 7mm (43
GHz) produced a result that offered a series of useful insights.
Specifically in \S 2, we describe our ATCA millimeter search, and in
\S 3, we describe similar millimeter work done with the VLA. In \S 4
we describe our combined results, and then in \S 5 we discuss the
implications of our null result.

\section{Observations}

We observed Gl~876 at both 3- and 7- mm with the Australia Telescope
Compact Array (ATCA) and the Very Large Array (VLA), respectively.
Table 1 summarizes the observing details.

\subsection{3-mm Australia Telescope Compact Array Observations}

Our 3mm ATCA observations were conducted on JD 2453626 with five of
ATCA's six 22-m antennas. Uranus was observed for 15 minutes as the
primary calibrator. The secondary calibrator (PKS B2246~+~121) was
then observed for 3 minutes and every fourth pair of
target/secondary scans were followed by a "paddle" observation for
absolute calibration. Total time on target was 2.5 hr with the IFs
set at 93.5 and 95 GHz. Table 1 provides further details on the
observations.

Data reduction was done using Multichannel Image Reconstruction,
Image Analysis and Display (Sault \& Killeen 1993, thereafter
MIRIAD). The data sets from both IFs were edited and calibrated
separately, but combined in imaging in order to maximize
sensitivity. From this we produced a continuum image with a
restoring beam of 2.75{\arcsec}$\times$2.75{\arcsec} with an RMS
noise floor at 0.9~$\pm$~0.01~mJy beam$^{-1}$. Nineteen beams were
the required minimum to cover the notional disk. We found no dust
emission in the image and stopped reduction there. Our image covered
the $\sim{55}\arcsec$ extent of a notional $\sim{200}$ AU debris
disk.

\subsection{7-mm Very Large Array Observations}

The 7mm observations at National Radio Astronomical Observatory
(NRA0) VLA took advantage of the dynamically scheduled period during
array reconfigurations. Epoch 1 was centered on JD 2453658 while
epoch 2 was centered on JD 2453676. Again, Table 1 details the
observations. For the first epoch, we used the hybrid DnC
configuration with the North arm in C configuration while the East
and West arms were in the more compact D configuration. The second
epoch occurred in full D configuration. For both of our VLA epochs,
we used the full complement of available antennas (23 of 27 25-m
antennas), and observed for a total 3.4 hours. The synthesized beam
of the VLA is 1.47$\arcsec$ $\times$ 0.93$\arcsec$. From the VLA
data we produced a 512 x 512 pixel image with a spacing of
$0.2\arcsec$ per pixel. The resulting $\sim$~100$\times$100$\arcsec$
image covers the 85$\arcsec$ extent of a 400~AU debris disk. We used
30 beams to cover the linear extent of the potential disk.

We reduced the two epochs using Astronomical Image Processing System
(Greisen 2006, thereafter AIPS) and calibrated each epoch
independently. We set the absolute flux density scale using a
computed flux density of $0.53$~Jy for the extragalactic calibrator,
3C48. Extragalactic calibrator source, PKS B2246~+~121, was used to
estimate the instrumental and atmospheric phase fluctuations. We
then applied phase corrections to the target source data. Gl~876 and
PKS B2246~+~121 were then imaged at each epoch and concatenated into
a single calibrated data set, and then imaged again. This resultant
35{\arcsec}$\times$35{\arcsec} contour image from the combined data
showed no apparent emission. The VLA dirty map's RMS noise floor was
at 44~$\pm$~2.5~$\mu$~Jy~beam$^{-1}$. Finally, we applied a variety
of \emph{uv}-plane tapers and ranges besides these to extract any
missed detection. This additional processing also failed to produce
a detection.

\section{Results}

In this section we use our upper limits on the millimeter emission
from Gl~876 to constrain the properties of any debris disk orbiting
it.  Following standard formulae (e.g., Lestrade et al. 2006, Dent
et al. 2006), we relate the flux density upper limits to the dust
mass of an optically thin debris disk as

\begin{equation}
S_{\lambda}~=~\frac{M_{dust}B(\lambda,~T_{d})\kappa}{{D}^2},\label{eq1}
\end{equation}

\noindent where $S_{\lambda}$ is the observed flux density at the
given wavelength, $M_{dust}$ is the dust mass in the disk, $\kappa$
is the mass opacity, and $B(\lambda, T)$ is the Planck blackbody
function for dust at temperature $T$. To provide a new constraint on
the dust mass, we solved for mass using a 3~$\sigma$ noise floor.

For the temperature $T$ of the dust particles, we assume a cool $T
\approx~20~K$ (see below), and also less radiant energy is
transferred to the dust than in the ideal case. Classic 1~$\mu$m
sized Lambertian, spherical dust particles are generally assumed to
have an albedo $a_d$ for the dust where $a_d~\approx~{0.06}$ (Brown,
et al., 1997; Jewitt, Luu \& Chen 1996; Luu \& Jewitt 1996).

Since the dependence of temperature with typical star-to-dust
distance is $d^{0.5}$, the temperature at 30~AU outward will differ
little, so we use one temperature to approximate the disk. But since
M dwarfs are less luminous than G type stares by $L~{\approx}~0.1$
to $0.001 L_\odot$, they irradiate their circumstellar dust to a
lesser $\leq~20~K$ than for the $35~to~50~K$ expected from dust
about solar-like stars (Lestrade et al. 2006; Beckwith et al. 1990).
The distance $D$ is known with high accuracy to be 4.69 pc.  For the
mass opacity in the most fundamental scenario, we presume the disk
to be optically thin, and thus adopt a standard value of

\begin{equation}
\kappa = {{\kappa}_0} \left( \frac {{ {\lambda}_0}}{{\lambda}}
\right) ^{\beta} \label{eq3}
\end{equation}

\noindent where ${\kappa}_0 = 1.0\,\mathrm{cm}^2 g^{-1}$. Owing to
the wide disparity in postulated values for the opacity spectral
index from 0.2 to 3.0, we will somewhat arbitrarily assume $\beta =
1$ as a starting point. Clearly our results will depend upon the
assumed temperature and mass opacity. Should debris disks around M
dwarfs turn out to have dust with, for instance, significantly lower
mass opacity than around earlier-type stars, we would have
underestimated the dust mass in the Gl~876 system.

To prepare for a notional detection, we first assumed a disk
diameter. At the distance of 4.69~pc, $1\arcsec$ equals 4.69~AU, so
the resolution (the number of divisions into which our telescopes'
beams are divided) is equivalent to a linear scale of 12.7~AU. One
possibility is that any disk would be unresolved because of its
size. The debris disk is unlikely to extend much closer in toward
the star than the outermost planet, which lies at a semi-major axis
of 0.2~AU. Such arrangement allows a scenario where any disk would
be unresolved with our instruments. However, guided by the known
debris disks around M dwarfs (e.g., AU Mic, Gl 842.2, Gl 182), we
shall assume a simplistic disk diameter of $\sim{200}$~AU (or
$\sim{42.6}~\arcsec$) here. This assumed disk would be well within
the primary beams of VLA and ATCA and would even allow a disk to be
resolved by the 4.69 divisions spread over the extent of the beam.
In fact, we made images of a much larger $>100\arcsec$ region as a
precaution, so that we could detect any disk as large as even
$225~AU$ in radius.

The assumed size of the disk becomes relevant to the value of the
noise floors of the VLA and the ATCA, which provide a limiting value
for observed intensity. These must be related to a flux density by
assuming the area of a possible disk.

It is important to understand how (and when) inclination affects any
detection above the noise floor.  First it is statistically unlikely
that the disk will be exactly face on, nor edge on. There exists a
higher probability of being detected at some intermediate
inclination. As with visual detections of edge-on transits at $i =
90^{\circ}$, the geometric a priori likelihood of a single
inclination is given by

\begin{equation}
\textsl{P} =0.0045\left( {\frac{1AU}{a}} \right)\left(
{\frac{R_\ast +R_{disk} }{R_\odot }} \right) , \label{eq4}
\end{equation}

\noindent where \textit{a} is the semi-major axis of the orbit,
$R_\ast$ is the radius of the star and $R_{disk}$ is the thickness
of the disk, arbitrarily chosen here to be $\sim3~R_{Jup}$. In the
case of Gl~876, we also assume $R_\ast$ = $0.3 R_\odot$, and for a
notional disk we choose \textit{a} to be a mean $100-AU$. The result
is $\sim{2}~\%$ for any given inclination about Gl~876, and a
strictly geometric probability of $\sim{40}~\%$ for a range of $i$
from $40^{\circ}$ to $60^{\circ}$. Probabilistically, it is more
likely that Gl 876 disk is not edge-on but has some lesser
inclination. The less the disk is inclined from `edge-on' to
`face-on', the more the disk surface brightness decreases, thus
making a null detection more likely as the flux density drops
beneath the noise floor.  As well, the disk must have some optical
thickness (as described for younger disks in Takeuchi \& Lin 2003,
2005) in order for us to have observed a gradient, and thus infer
any tilt. As the study of M star disks is relatively adolescent, it
is not clear at this point how thick Gl 876's disk might be. In any
event, such an opportunity for a disk to escape detection would be
consistent with previous optical observations.

We solved for dust mass using the RMS noise floor for each radio
telescope as a threshold, or minimum detectable mass. The 3~$\sigma$
upper limit of 135 $\mu$Jy on any undetected mass then becomes
0.0006 M$_{\oplus}$ for the area of a nominal $200~AU$ radius disk,
for the more stringent VLA results.

\section{Discussion}

As previously mentioned, Trilling, Brown, \& Rivkin (2000) used
NASA's Infrared Telescope Facility (IRTF) Cold Coronagraph (CoCo) at
1.62 $\mu$m to search for a circumstellar disk around Gl~876, and
produced their own null result for this system. However, their
observations were less sensitive (3.6 times less so), and moreover
were restricted to a narrow, $5\arcsec$ (25-AU) beam width. Based on
the size of the few red dwarf disks detected since their
observations, this beam width was likely insufficient to assert that
any M dwarf disk does or does not exist about Gl~876. This beam
easily could have missed a disk by looking at the cleared cental
hole in it.

Other observations of nearby stars done by Greaves et al. (2004) in
fact provided additional impetus to do similar observations for
Gl~876. From their observations they estimate that the
$\epsilon$~Eridani and $\tau$ Ceti dust disks have $0.016$ and
$0.0005~M_{\oplus}$ dust masses, respectively. As further
comparison, AU~Mic is a much younger M star at double the distance
($\sim{10}$~AU), and whose edge-on disk has a radius is between 50
and 210 AU. While the AU~Mic disk is just one example, its minimum
size also gave us confidence that any Gl~876 disk would also likely
be resolved in our VLA and ATCA observations. Admittedly, strict
comparisons between AU~Mic and Gl~876 would be limited owing to the
age difference, but seeing an older disk about Gl~876 would allow
this age difference to be exploited in a first-time age comparison.

In the end, our improved beam width and sensitivity still proved
insufficient to detect a larger, fainter disk similar to these, but
we can say that our observations had sufficient beam width to surely
detect any Kuiper-like disk whose flux rose above the VLA noise
floor of $44~{\mu}$Jy, \emph{if} other factors did not cause the
non-detection. In such a simplistic scenario we assert that no disk
exists to the mass limit posed, if we assume that the dust is
essentially optically thin.

The mass constraint that an upper mass limit puts on Gl~876 has
dynamical implications worth exploring. The reasoning which allows
resolved disks to be used to infer an inclination follows Lestrade
et al. (2006). As noted earlier, they resolved a debris disk about
GJ~842.2 at a shorter wavelength of 0.85~mm. This was done with
sufficient resolution on James Clerk Maxwell Telescope's
Submillimetre Common-User Bolometer Array (SCUBA). Their results
suggest that GJ~842.2 was generally inclined, for the same reasons
inclination could be discernable for Gl~876.

The curious dynamics the Gl~876 system exhibits is due to its two
outer Jovian planets which are locked in a 2:1 mean resonance,
discussed in Laughlin et al. (2005), Laughlin, Bodenheimer \& Adams
(2004), and Marcy et al. (2001). In 2002, Benedict et al. reported
that their \emph{Hubble} FGS astrometry of Gl~876 revealed an
inclination of $i \approx 90^{\circ}$. However, the 2005 inner
planet detection prompted Rivera et al. to revisit the inclination
issue, which instead appeared to be $i \approx 50^{\circ}$. This
more-tilted inclination was found to be consistent with $3\sigma$
photometry and radial velocity transit reduction by one of us
(Shankland et al. 2006), using reduction methods shown viable in
Kane (2007). It is statistically likely that a positive dust
detection would have probably suggested some inclination if there
were some opacity -- and thus support of one set of these
conflicting optical observations.

On the other hand, a negative detection (as is the result here)
suggests that a thin dust mass could still exist about Gl~876
\emph{if} the dust density were below the detection threshold of the
\emph{individual} VLA or ATCA resolution-per-pixel, \emph{or} the
overall upper mass limit. The yet-poorly understood dust density,
temperature, spectral index, opacity and optical thinness muddies
our understanding of the mechanisms at play, and inclination would
be a factor in each of these. Still, there are other possibilities
for our null result. Another may be that the formation and evolution
of systems (disks and planets) is different in M dwarfs. At the
least, any unexpanded composition in the system would lead to
misunderstood opacities, albedos, radii, or black body behavior.

As we have suggested, we also could have missed any disk owing to a
less-than edge-on orientation that would have reduced the surface
brightness in a less-than transparent disk. Slipping under this
threshold (and the noise floor of VLA and ATCA) at lower
inclinations would be a result which favors the $i \approx
50^{\circ}$ scenario posited by Rivera et al. (2005) and Shankland
et al. (2006). Benedict et al.'s $i \approx 90^{\circ}$ could
instead be correct if the disk were edge on, but the mass is small,
at 0.0006~M$_\oplus$. Our non-detection thus offers a two-variable
constraint. Until the basis for a non-detection is constrained
further, we can also at best \emph{suggest} that Greaves' hypothesis
appears to remain intact. More work clearly needs to be done to
understand the properties of debris dust about M stars. Further, if
these initial suppositions are correct, Gl~876's lack of dust also
suggests few planetesimals in the system.

So our results suggest further scrutiny of M stars is needed in
order to understand how their systems differ from solar-type stars,
particularly since M dwarfs have a demographic monopoly on the
galaxy. More sensitive observations of this and other M stars would
also begin a foundation for further numerical modeling of their
disks (or lack thereof), as suggested in Deller \& Maddison (2005).
Certainly, any connection of dust to terrestrially-massed planets
will fuel an interest in the increased scrutiny of M dwarfs. From
the recent optical detections, the growing consensus is that red
dwarfs may very well harbor the first discovered exo-Earths.

It is also worth mentioning that because of the low luminosity of M
dwarfs (below $0.1~L_{\odot}$) that leads to cooler dust about them,
sub-millimeter or millimeter telescopes may be more sensitive than
mid-infrared ones in detecting such planet-associated-disks. We
would encourage observations at these wavelengths be explored
further. In particular, we recommend that the six M~dwarfs found
with planets so far be comprehensively checked for dust, to include
Gl~876 (with greater sensitivity than us), Gl~436 (Butler et al.
2004), Gl~674 (Bonfils et al. 2007), Gl~849 (Butler et al. 2006),
Gl~581 (Udry et al. 2007), and GJ~317 (Johnson et al. 2007).
Understanding the dust in such planet-bearing systems (as addressed
in Dutrey, des~Etangs, \& Augereau 2004) may be a key not just to
making a first exo-Earth detection, but will more importantly offer
a broader understanding of planets and their formation about the
populous M-type stars.

\section{Conclusions}

We used the VLA and ATCA at millimeter wavelengths to search for
thermal radiation from cool dust from the GL~876 system, and
achieved the null result which improved upon previous limits. We
observed no such emissions to 3$\sigma$, at the 135 $\mu$~Jy RMS
detection threshold. From this we calculated that any dust mass that
might still be there had to be constrained to be a mass less than
0.0006 M$_{\oplus}$ for a nominal 200~AU radius disk. Our $3-\sigma$
noise floor was established during our most sensitive observations
with the VLA, and were consistent with our ATCA observations. This
constraint does not generally contravene Greaves' postulation that a
system's bodies sweep out regenerated dust. Since the disk
inclination affects the surface brightness (depending on our
`emerging' understanding of opacity, temperature, spectral index,
density and thinness), this lent qualitative constraints on how the
system might be inclined. All things being equal a non-detection is
more consistent with a lower inclination than a higher one.

For Gl~876, our result most importantly places basic constraints on
the limits afforded by the instrumentation, and a more stringent
upper M$_{\oplus}$ limit on any potential exo-debris there. While a
lower mass debris belt \emph{might} be found for Gl~876 with greater
sensitivity, the null observations we find thus far corroborate the
effects of close-in Jovian planets. For Gl~876 in particular, we
offer four solutions to explain this non-detection. Alternatively
the system is less dusty or is optically thinner at the more
detectable $i \sim{90}^{\circ}$, or is `dustier' yet less detectable
at some less edge-on inclination. The third possibility is that this
red dwarf dust is comprised of material which is not similar to
comparable disks about solar-like stars (e.g., has a different
spectral index, thinness, temperature or opacity), and so evades
detection for now. Finally, while it would be physically very
unlikely, a disk (or instead a central hole) could have been larger
than $\sim$~400~AU. Answers will remain enigmatic until a more
sensitive dust study of Gl~876 is done, and more generally, a
robust, low-bias census of M star systems is completed.

\vspace{\bigskipamount}
\vspace{\bigskipamount}

We thank K. Johnston, G. Laughlin, R. Gaume, and J. Pier for useful
discussions, feedback or other support. We also thank referee, whose
insightful feedback greatly improved the presentation of this paper.
The Australia Telescope is funded by the Commonwealth of Australia
for operation as a National Facility managed by CSIRO. The National
Radio Astronomy Observatory is a facility of the National Science
Foundation operated under cooperative agreement by Associated
Universities, Inc. NRL astronomical research is supported by Navy
6.1 funding, and USNO astronomy is supported by Navy RTDE and OMN
funding. PDS thanks the NRAO staff for their support at Socorro. DLB
thanks the staff of the ATNF for their hospitality both at Narrabri
and Epping.



\clearpage

\normalsize
\clearpage
\begin{center}
\textbf{Table 1: ATCA and VLA Configurations}

\end{center}
\begin{table}[htbp]
\begin{center}
\label{tab1Obs}
\begin{tabular}{l||l|l}
\hline\hline
Parameter&ATCA&VLA \\
\hline
Date ($2005$)& 12 Sep & 14 Oct (Epoch 1) \\

Date ($2005$)& -- & 01 Nov (Epoch 2) \\

Julian Date & JD2453626  & JD2453658 (Epoch 1) \\

Julian Date & -- & JD2453676 (Epoch 2) \\

Frequency ($GHz$) & 93.5, 95 ($\sim{3}$~mm) & 43.315, 43.365 ($\sim{7}$~mm) \\

Mode & Continuum & Continuum \\

Elements & 5 of 6 & 23 of 27  \\

Bandwidth ($MHz$) & 128, 128 & 50 \\

Synthesized Beam ($arcsec^{2}$) & 2.75$\times$2.75 & 1.47$\times$0.93  \\

Configuration & H168 & DnC, D \\

Time on Target ($hr$) & 2.5  & 3.4 \\

Cycle ($sec$) & 10 & 80 (source), 30 (cal) \\

Epochs Total ($hr$) & 1~x~4.5 & 2~x~2 \\

$S_{3\sigma}$ ($mJy~beam^{-1}$) & 0.9~$\pm~0.01$~ & 0.04$\pm$~0.003  \\

Phase Calibrator & PKS B2246~+~121 & PKS B2246~+~121 \\

Flux density calibrator & Uranus & 3C48 \\

\hline\hline
\end{tabular}
\end{center}
\end{table}


\begin{thebibliography}{DUM}

\bibitem[]{}
Apai, D., Luhman, K., \& Liu, M. 2007, ArXiv e-prints, arXiv:
0702286v1

\bibitem[]{}
Beckwith, S., Sargent, A., Shini, R., \& G\"{u}sten, R. 1990, AJ,
99, 924

\bibitem[]{}
Benedict, G. F.; et al. 2002, ApJL 581, 115

\bibitem[]{}
Bonfils, X., et al. 2007, ArXiv e-prints, 704, arXiv:0704.0270

\bibitem[]{}
Brown, R., Cruikshank, D., Pendleton, Y., \& Veeder, G. 1997,
Science, 276, 937

\bibitem[]{}
Butler, R.~P., Johnson, J., Marcy, G., Wright, J., Vogt, S., \&
Fischer, D.~A. 2006, \pasp, 118, 1685

\bibitem[]{}
Butler, R., et al. 2004. ApJ, 617, 580

\bibitem[]
Carrol, B., \& Ostlie, D. 2007, An Introduction to Modern
Astrophysics (2nd Ed.; San Francisco: Pearson Addison Wesley)

\bibitem[]{}
Delfosse X, et al. 1998. A\&A, 338, L67

\bibitem[]{}
Deller, A., Maddison, S. 2005, ApJ, 625, 398

\bibitem[]{}
Dent, W., Walker, H., Holland, W., \& Greaves, J. 2000, 314, 702

\bibitem[]{}
Dutrey, A., des Etangs, A., Augereau, J. 2004, in Space Science
Ser., COMETS II, ed. R. Binzel (Tuscon: U of AZ), in
astro-ph/0404191v1

\bibitem[]{}
Gautier, T., et al. 2007, ApJ, 667, 527

\bibitem[]{}
Greaves, J., Holland, W., Jayawardhana, R., Wyatt, M., Dent, W.
2004, MNRAS, 348, 1097

\bibitem[]{}
Greisen, E. et al. 2006, AIPS Cookbook, http://www.aips.nrao.edu/cook.html

\bibitem[]{}
Jewitt, D., Luu, J., \& Chen, J. 1996, AJ, 112, 1225

\bibitem[]{}
Johnson, J., Butler, R., Marcy, G., Fischer, D., Vogt, S., Wright,
J., \& Peek, K. 2007, ArXiv e-prints, 707, arXiv:0707.2409

\bibitem[]{}
Kalas, P., Liu, M., \& Matthews, B. 2004, Science, 303, 1990

\bibitem[]{}
Kane, S. 2007, MNRAS, 380, 1488

\bibitem[]{}
Laughlin, G., Bodenheimer, P., \& Adams F. C. 2004, ApJ, 612, L73

\bibitem[]{}
Laughlin, G., Butler, R., Fischer, D., Marcy, G., , Vogt, S., Wolf,
A. 2005, ApJ, 622, 1182.

\bibitem[]{}
Lestrade, J., et al. 2006, A\&A, 460, 733

\bibitem[]{}
Lissauer, J. ApJ, 2007, 660, L149

\bibitem[]{}
Liu, M. C., et al. 2004, ApJ, 608, 526

\bibitem[]{}
Luu, J., \& Jewitt, D. 1996, AJ, 111, 499

\bibitem[]{}
Marcy, G. W., Butler, R. P., Fischer, D., Vogt, S. S., Lissauer, J.
J., \& Rivera, E. J. 2001, ApJ, 556, 296

\bibitem[]{}
Marcy, G., Butler, R., Vogt, S., Fischer, D., Lissauer, J. 1998, ApJ,
505, L147

\bibitem[]{}
Perryman, M., et al. 1997, AA, 323, 49

\bibitem[]{}
Quillen, P. C. \& Thorndike, S. 2002, ApJ 578, L149

\bibitem[]{}
Rivera, E. J., et al. 2005, ApJ, 634, 625

\bibitem[]{} V. Roccatagliata, V., Henning, Th., Wolf, S., Carpenter, J. M., \&
Rodmann, J. 2007, in IAU Symp. 243, Star-disk Interaction in Young
Stars (Grenoble: IAU), 21-25 May,
http://www.iaus243.org/IMG/pdf/Roccatagliata.pdf

\bibitem[]{}
Sault, R. \& Killeen, N. 1993, MIRIAD Cookbook,
http://www.atnf.csiro.au/ computing/software/miriad/

\bibitem[]{}
Shankland, P., et al. 2006, ApJ, 653, 700

\bibitem[]{}
Takeuchi, T., \& Lin, D. 2003, ApJ, 593, 524

\bibitem[]{}
Takeuchi, T., \& Lin, D. 2005, ApJ, 623, 482

\bibitem[]{}
Th$\acute{e}$bault, J., Augereau, J., Beust, H. 2003, AA, 408, 775

\bibitem[]{}
Trilling, D. E., Brown, R. H., \& Rivkin, A. S. 2000, ApJ, 529, 499

\bibitem[]{}
Udry, S., et al. 2007, A\&A, 469, L43

\bibitem[]{}
Wyatt, M. C., et al. 1999, ApJ, 527, 918


\end{thebibliography}
\end{document}